\documentstyle[preprint,aps]{revtex}
\newcommand{\krc}{k r_c}
\newcommand{\krcp}{k r_c\pi}

\begin{document}

\setlength\baselineskip{20pt}

\preprint{\tighten\vbox{\hbox{CALT-68-2286}\hbox{hep-th/0007065}}}

\title{Quantum Stabilization of Compactified $\mbox{AdS}_5$}

\author{Walter D. Goldberger$^a$\footnote{walter@theory.caltech.edu}
and Ira Z. Rothstein$^{a,b}$\footnote{ira@cmuhep2.phys.cmu.edu}}
\address{
\vspace{.5cm} 
$^a$California Institute of Technology, Pasadena, CA 91125\\
$^b$Department of Physics, Carnegie Mellon University, Pittsburgh, PA 15213}

\maketitle

{\tighten
\begin{abstract}
We investigate the role of quantum fluctuations in the system composed
of two branes bounding a region of AdS.  It is shown that the modulus
effective potential generated by quantum fluctuations of both brane
and bulk fields is incapable of stabilizing the space naturally at the
separation needed to generate the hierarchy.  Consequently, a
classical stabilization mechanism is required.  We describe the
proper method of regulating the loop integrals  
and show  that, for large brane separation,  the quantum effects are power
suppressed and 
therefore have negligible affects on the bulk dynamics once a classical
stabilization mechanism is in place.

\end{abstract}}
\vspace{0.7in}
\narrowtext

\newpage

Recently, Randall and Sundrum~\cite{RS} proposed a novel mechanism for
addressing the hierarchy problem.  In their model, the Standard Model
fields are confined to one of two 3-branes which are endpoints of an $S^1/Z_2$ orbifold spatial dimension.  A negative bulk
cosmological constant generates an AdS metric in the five-dimensional
spacetime
\begin{equation}
\label{eq:metric}
ds^2 = e^{-2 k r_c|\phi|}\eta_{\mu\nu} dx^\mu dx^\nu - r_c^2 d\phi^2,
\end{equation}
where $k$ is a parameter of order the Planck scale which is related to the AdS
radius of curvature and $r_c$ determines the length of the orbifold.
The coordinate $\phi\in [-\pi,\pi]$ parameterizes the fifth dimension,
with the point $(x,\phi)$ and $(x,-\phi)$ identified, and the two
3-branes reside at the orbifold fixed points $\phi=0,\pi$.  Because of
the exponential factor in Eq.~(\ref{eq:metric}), a field with
Lagrangian mass parameter $m_0$ that is confined to the brane at
$\phi=\pi$ (the ``TeV brane'') will have a physical mass $m=m_0
e^{-\krcp}.$ If all mass scales in the theory are of order the Planck
scale and $\krc\sim 12,$ then the observed mass $m$ is in the TeV
range.  In this way, the hierarchy between the TeV and the Planck
scale is generated purely through gravitational effects.

The Randall-Sundrum (RS) scenario contains a modulus field that determines the size of
the $S^1/Z_2$ orbifold extra dimension.  This scalar arises as one of
the massless fluctuations about the background AdS geometry, and it is
encoded in the five-dimensional metric as
\begin{equation}
\label{eq:fluc}
ds^2 = e^{- 2 k |\phi| T(x)} g_{\mu\nu}(x) dx^\mu dx^\nu - T^2(x) d\phi^2,
\end{equation}
where the field $g_{\mu\nu}$ is the four-dimensional graviton and
$T(x)$ is the modulus, or ``radion'' field whose VEV $r_c=\langle
T\rangle$ determines the length of the orbifold according to
Eq.~(\ref{eq:metric}).  Dimensional reduction of the five-dimensional Einstein-Hilbert action for Eq.~(\ref{eq:fluc}) leads to an effective action for the
massless fields~\cite{GW2,csaki}
\begin{equation}
\label{eq:action}
S= \frac{2 M^3}{k} \int d^4 x \sqrt{-g}{}\left(1- (\varphi/f)^2\right)
R +\frac{1}{2}\int d^4 x\sqrt{-g} {}\partial_\mu \varphi \partial^\mu
\varphi,
\end{equation}
where $R$ is the Ricci scalar constructed from $g_{\mu\nu}$ and we
have defined $\varphi = f \exp(-k\pi T)$ with $f=\sqrt{24 M^3/k}.$

To account for the observed discrepancy between the gravitational and
electroweak scales, we need $kr_c\sim 12$.  However, there is nothing
in Eq.~(\ref{eq:action}) that stabilizes the VEV of the radion
$T$\footnote{In fact, the original RS solution necessitates one fine
tuning of parameters to keep the radion potential flat and the
resulting metric static.  This fine tuning problem goes away in the
presence of some additional radion stabilization dynamics.  There is a
second fine tuning, related to the cosmological constant problem, about which the RS solution has nothing to say.}.  Some
additional dynamics must be introduced to make $kr_c\sim 12$ without
fine tuning parameters.  As suggested in~\cite{GW1}, introducing a
bulk scalar with appropriate interaction terms on the branes can
induce a radion potential that has an acceptable minimum without severe
tuning of the model parameters.  Although in~\cite{GW1} the scalar
profile was treated as a perturbation on the background metric, its
back reaction on the spacetime geometry can be included~\cite{deWolfe}
without changing the qualitative features of the stabilization
mechanism.

The analysis of refs.~\cite{GW1,deWolfe} was purely at the classical
level.  However, quantum fluctuations of fields which propagate in the
bulk or on the TeV brane will also generate contributions to the
effective radion potential\footnote{Fields on the Planck brane, at
$\phi=0,$ do not couple directly to the radion.  We expect that their
contribution to the one-loop effective potential is suppressed
relative to the sources mentioned above.}. In this paper we explore
the possibility that it is these quantum corrections which stabilize
the radion. We calculate the effective potential arising from bulk
fields as well as fields confined to the TeV brane. For the confined
fields we calculate using three different regulators, and show clearly
that the effective cutoff on the brane is indeed of order TeV.  After proper regularization, the sole effect of the brane field fluctuations is the renormalization of the brane tension.  The physical contribution to the effective potential from integrating out bulk fields is suppressed by large powers of
$\exp{(-kr_c\pi)}$ in the large $r_c$ limit.  As a consequence, the resulting
vacuum energy cannot stabilize the brane separation naturally. Furthermore, the
quantum effects do not spoil the classical mechanism of~\cite{GW1}.
Our results resolve a discrepancy between two previous results in the
literature\cite{garriga,toms}.

\section{Quantum Corrections to the Radion Potential}
\label{sec:quantum}

First, we consider the contribution to the radion potential coming from a field
on the TeV brane.  We will show that the fluctuations of fields on the brane serve only to renormalize the brane tension. Given the subtlety in the regularization procedure we will calculate using three different regulators.

For concreteness, we will take a scalar field theory confined to the
TeV brane.  First we shall compute the effective potential using
dimensional regularization.  In $n=4-\epsilon$ dimensions, the action
is given by
\begin{equation}
S={1\over 2}\int d^n x a^n\left({1\over a^2}(\partial h)^2 - m_0^2 h^2\right),
\end{equation}
where $m_0$ is of order the Planck scale, and the powers
of $a$ multiplying the kinetic and mass terms come from the induced
metric on the brane.  To compute the radion potential, take $a$
constant and rescale $h\rightarrow a h$.  The rescaled field has a
canonically normalized kinetic term and an effective mass $m=a m_0.$
The effective potential obtained from integrating out $h$ can be
trivially expressed as the zero point energy in the presence of a
constant $\varphi$ field configuration:
\begin{equation}
V = {1\over 2} \mu^{4-n} \int {d^{n-1} k\over
(2\pi)^{n-1}}\sqrt{k^2 + a^2 m_0^2},
\end{equation}
with $\mu$ an arbitrary mass scale which has been introduced to keep
$V$ a four-dimensional energy density.  The resulting expression,
\begin{equation}
V=-{1\over 2}{\mu^{4-n}\over (4\pi)^{n/2}}\Gamma\left(-{n\over 2}\right) (m_0 a)^n,
\end{equation}
contains a divergent piece that must be absorbed into a local
counterterm.  Such a counterterm is provided by the brane tension on
the TeV brane
\begin{equation}
S_{ct} = -\int d^n x a^n\delta V \mu^{n-4},
\end{equation}
which is generally covariant in $n$ dimensions.  Comparing this with
our result, we see that $V$ is in fact pure counterterm: the effect of
the scalar $h$ is simply to renormalize the brane tension. Given
a bare mass of order the Planck scale (this is the appropriate choice
for our set of coordinates), there are no large logs for $\mu\simeq M_{pl}$.

An alternative way of understanding this result is to regulate the
divergent integral using a physical (coordinate invariant) cutoff
$\Lambda$.  The vacuum energy for $h$ is then, for $\Lambda\gg m_0$
\begin{eqnarray}
\nonumber V &=&{1\over 2}\int^{\Lambda a} {d^3 k\over (2\pi)^3}\sqrt{k^2+a^2
m_0^2}\\ &=& {a^4\over 32\pi^2}\left[2\Lambda ^4 +\Lambda^2 m_0^2 +
m_0^4\ln\left({m_0\over 2\Lambda}\right)\right],
\end{eqnarray}
which simply induces a shift in the brane tension.  Note that the
coordinate cutoff on the momentum integral is rescaled by a factor of
$a$ with respect to the physical cutoff.  Had we used an
$a$-independent cutoff on the momentum integral, we would have
generated terms in the effective action proportional to $(\Lambda m_0
a)^2.$  On the other hand, the
rescaled cutoff yields results that are consistent with
four-dimensional general covariance on the brane, and which are in
agreement with dimensional regularization. 

The same conclusion can be reached by using a Pauli-Villars regulator.
 To get a consistent result, the regulator fields must couple to the
 induced metric on the TeV brane in the same way as our scalar field.
 Performing the calculations in two dimensions for simplicity, we make
 the subtraction
\begin{equation}
V\rightarrow V-{1\over2(M_1^2-M_2^2)}\int {d^2 k\over (2\pi)^2}\left[(m^2-M_2^2)\sqrt{k^2+M_1^2}+(M_1^2-m^2)\sqrt{k^2+M_2}\right].
\end{equation}
Where $all$ the masses, including the regulator masses get rescaled by the warp factor.  Performing the momentum integral, it is easily seen that all logarithmic dependence on $a$ cancels from the regulated expression.  The remaining dependence on $a$ is a pure counterterm.

The quantum fluctuations of bulk fields also contribute to the radion
effective potential.  Decomposing the bulk field into four-dimensional
Kaluza-Klein modes, the potential can again be expressed as a sum over
zero point energies
\begin{equation}
\label{eq:kkpot}
V = (-1)^F {g\over 2}\mu^\epsilon \sum_n\int {d^{3-\epsilon} k\over
(2\pi)^{3-\epsilon}}\sqrt{k^2 + m_n^2},
\end{equation}
where $F=0,1$ for bosons and fermions respectively, and $g$ is the number of physical polarizations
of the Kaluza-Klein modes.  In this equation, the dependence on $a$
enters through the Kaluza-Klein masses $m_n.$ Defining $m_n=a k x_n,$
the above becomes
\begin{equation}
\label{eq:qbulk}
V= (-1)^{F+1} g {k^4 a^4\over 32\pi^2} \left({k^2 a^2\over
4\pi\mu^2}\right)^{-\epsilon/2}\Gamma(-2+\epsilon/2)\sum_n
x_n^{4-\epsilon}.
\end{equation}

We now evaluate Eq.~(\ref{eq:qbulk}) for a bulk scalar field with action
\begin{equation}
S_b={1\over 2}\int d^4 x\int_{-\pi}^\pi d\phi \sqrt{G}
\left(G^{AB}\partial_A \Phi \partial_B \Phi - \left(m^2 +\alpha{\sigma''\over
r_c^2}\right)\Phi^2\right),
\end{equation}
where $G_{AB}$ with $A,B=\mu,\phi$ is given by Eq.~(\ref{eq:metric}).  Because 
$\sigma'' = 2 k r_c\left[\delta(\phi)-\delta(\phi-\pi)\right],$ the parameter $\alpha$ controls a possible mass term on the boundaries
of the space.  Such mass terms arise if the field $\Phi$ is a
component of a supermultiplet on $\mbox{AdS}_5$ with one dimension
compactified on an $S^1/Z_2$ orbifold (see~\cite{GP}).  It is found in~\cite{GP} that the roots $x_n$ satisfy
\begin{equation}
j_\nu (x_n) y_\nu (a x_n) - j_\nu (a x_n) y_\nu (x_n)=0,
\end{equation}
where $\nu=\sqrt{4+m^2/k^2}$, $j_\nu (z)= (2-\alpha) J_\nu (z) + z
J_\nu'(z),$ and $y_\nu$ is given by the same expression with $Y_\nu$
replacing $J_\nu.$ (See~\cite{GP,KK} for other work on the Kaluza-Klein reduction of bulk fields.)  The $a$ dependence from the sum over $x_n$ in
Eq.~(\ref{eq:qbulk}), can be calculated by zeta function
regularization techniques~\cite{romeo}, which we now review.  

First, convert the sum into a contour integral
\begin{equation}
\label{eq:contour}
\sum_n x_n^{-s}={s\over 2\pi i}\int_C dz z^{-s-1}\ln{\left[j_\nu (z) y_\nu (a z) - j_\nu (az) y_\nu (z)\right]},
\end{equation}
which is valid for $\mbox{Re}s>1$. In this equation, $C$ is a contour
between arcs of radius $\delta$ (chosen to avoid a possible pole at
$z=0$) and $R\rightarrow\infty$ which circles the roots $x_n$ in a
counterclockwise manner.  Our goal is to perform the analytic
continuation of the RHS of Eq.~(\ref{eq:contour}) to a neighborhood of
$s=-4.$ To do this, split the contour into $C_+$ and $C_-,$ its
portions above and below the real axis respectively.  On each contour,
the asymptotic expansion of the argument of the logarithm is
\begin{equation}
Z_\nu(z,a)\equiv j_\nu (z) y_\nu (a z) - j_\nu (az) y_\nu (z)\sim
\mp{i\over\pi} z\sqrt{a}e^{\mp iz(1-a)}\left[1+{\cal
O}\left(1/z\right)\right].
\end{equation}  
We now add and subtract the logarithm of the RHS of this expression to the contour integral above, which yields
\begin{eqnarray}
\label{eq:splitint}
\nonumber \sum_n x_n^{-s}&=&{s\over 2\pi i}\sum_{C_\pm}\int_{C_\pm} dz
z^{-s-1}\ln\left[\pm{i\pi\over z\sqrt{a}}e^{\pm iz(1-a)}
Z_\nu(z,a)\right] \\ &&{}-{s\over 2\pi i}\sum_{C_\pm}\int_{C_\pm} dz
z^{-s-1}\ln\left[\pm{i\pi\over z\sqrt{a}}e^{\pm iz(1-a)}\right].
\end{eqnarray}
The first line is now defined for $\mbox{Re}s>-1,$ while the second is
still only defined for $\mbox{Re}s>1.$ However, for the second term in
Eq.~(\ref{eq:splitint}), we are free to deform the contour $C$ into a
straight line running parallel to the imaginary axis from
$z=i\infty+\delta$ to $z=-i\infty+\delta.$ The result is
\begin{eqnarray}
\nonumber \sum_n x_n^{-s}&=&{s\over 2\pi i}\sum_{C_\pm}\int_{C_\pm} dz
z^{-s-1}\ln\left[\pm{i\pi\over z\sqrt{a}}e^{\pm iz(1-a)}
Z_\nu(z,a)\right]\\ &&{}-{s\over\pi}\left[{2(1-a)\over
(1-s)}\delta^{1-s}+{\pi\over 2 s}\delta^{-s}\right].
\end{eqnarray}
Since the second line of this equation provides its own analytic
continuation, we can now extend the definition of the sum on the LHS
to $-1<\mbox{Re}s<0.$ In this region, it is safe to take the limit
$\delta\rightarrow 0.$  Then second term above vanishes.  
To evaluate the piece left over, we can take the straight
line contour along the imaginary axis.  The result, valid for
$-1<\mbox{Re}s<0,$ is
\begin{equation}
\label{eq:intrep}
\sum_n x_n^{-s}={s\over\pi} \sin\left({\pi s\over
2}\right)\int_0^\infty dt\,t^{-s-1}\ln\left[{2\over t\sqrt{a}}
e^{-t(1-a)}\left\{k_\nu (t) i_\nu (a t) - k_\nu (at) i_\nu
(t)\right\}\right],
\end{equation}
where $i_\nu (t)=(2-\alpha) I_\nu (t)+t I_\nu'(t),$ and $k_\nu(t)$ is
defined in the same way with $K_\nu(t)$ instead of $I_\nu(t).$

Eq.~(\ref{eq:intrep}) still needs to be extended to a neighborhood of
$s=-4.$ For $s=-4+\epsilon,$ it can be written as
\begin{eqnarray}
\label{eq:adep}
\nonumber \sum_n x_n^{-s}&=& -2\epsilon\Bigg\{\int_0^\infty
dt\,t^{3+\epsilon}\ln\left[1-{k_\nu(t) i_\nu(at)\over k_\nu(a t)
i_\nu(t)}\right]\\ &&{} + \int_0^\infty dt
\,t^{3+\epsilon}\ln\left[\sqrt{{2\pi\over t}} e^{-t}
i_\nu(t)\right]+{1\over a^{4-\epsilon}}\int_0^\infty
dt\,t^{3+\epsilon}\ln\left[-\sqrt{{2\over \pi t}} e^{t}
k_\nu(t)\right]\Bigg\}.
\end{eqnarray}
Because of the overall factor of $a^{4-\epsilon}$ in Eq.~(\ref{eq:qbulk}), the second two terms in this expression yield contributions that go as $a^{4-\epsilon}$ or independent of $a$ respectively.  The term that is independent of $a$ can be absorbed into the renormalization of the Planck brane tension.  As we discussed in the case of a TeV brane fields, $a^{4-\epsilon}$ can also be cancelled by a local counterterm.  The first term in the brackets is well defined at $s=-4.$  Taking the limit $\epsilon\rightarrow 0,$ we end up with
\begin{equation}
\label{eq:scalar}
V=V_h + V_v a^4 + {k^4 a^4\over 16\pi^2}\int_0^\infty dt\,t^3 \ln\left[1-{k_\nu(t) i_\nu(at)\over k_\nu(a t) i_\nu(t)}\right],
\end{equation}
where $V_{h,v}$ are shifts in the brane tensions.  For $a\ll 1,$ the $a$ dependence in the above equation is 
\begin{equation}
\label{eq:subleading}
\int_0^\infty dt\,t^3 \ln\left[1-{k_\nu(t) i_\nu(at)\over k_\nu(a t)
i_\nu(t)}\right]={2\over\nu\Gamma(\nu)^2}\left({\nu-\alpha+2\over
\alpha+\nu-2}\right)\left({a\over 2}\right)^{2\nu}\int_0^\infty
dt\,t^{2\nu+3} {k_\nu (t)\over i_\nu(t)}+{\cal O}(a^{2\nu+2})
\end{equation}
if $\alpha+\nu\neq 2.$  For $\alpha+\nu=2$ 
\begin{equation}
\int_0^\infty dt\,t^3 \ln\left[1-{k_\nu(t) i_\nu(at)\over k_\nu(a t)
i_\nu(t)}\right]={2(\nu-1)\over\Gamma(\nu)^2}\left({a\over
2}\right)^{2\nu-2}\int_0^\infty dt\,t^{2\nu+1} {k_\nu (t)\over
i_\nu(t)}+\dots,
\end{equation}
with terms of order $a^{2\nu}$ for $\nu\neq 2$ and $a^4\ln a$ for
$\nu=2$ not shown.  Incidentally, the eigenvalues $x_n$ for bulk
fields of higher integer spin satisfy equations that are identical to
that of the bulk scalar except for the values of $\nu$ and
$\alpha$~\cite{GP}.  It follows immediately that in those cases the
$a$ dependence is similar to that in Eq.~(\ref{eq:scalar}).
Furthermore, we can use the scalar result to calculate the effective
potential in these cases as well.  For instance, the contribution from
a bulk $U(1)$ gauge field can be obtained by taking $\nu=1$ and
$\alpha=1.$  To calculate the effective potential due to metric fluctuations, decompose the metric into a sum of four-dimensional scalar, vector, and transverse traceless modes.  The metric scalar and vector contributions are as described above, while the
 transverse traceless piece generates a term like Eq.~(\ref{eq:scalar}) with $\nu=2,\alpha=0.$

In addition to contributions from fields in the bulk and on the TeV
brane, the vacuum energy receives corrections from loops of the radion
itself.  These can be computed from the effective four-dimensional
Lagrangian,
\begin{equation}
{\cal L} = {f^2\over 2} (\partial a)^2 -\delta V_v a^4,
\end{equation}
where $\delta V_v$ is a small classical shift in the TeV brane tension
relative to the value which generates the background metric.  The
one-loop effective potential generated by the radion is
\begin{eqnarray}
\nonumber V &=&{1\over 2}\int^{\Lambda a} {d^3 k\over (2\pi)^3}\sqrt{k^2+{\hat
m}^2 a^2}\\ &=& {a^4\over 32\pi^2} \left[2\Lambda^4 + \Lambda^2 {\hat
m}^2 +{\hat m}^4\ln\left({\hat m}\over 2 \Lambda\right)\right],
\end{eqnarray}
where ${\hat m}^2 = 12\delta V_v/f^2$.  Note that as in the case of
TeV brane fields, we have used an $a$-dependent cutoff on the momentum
integral.  It is not immediately clear that
this is the correct cutoff to use in the dimensionally reduced theory,
which provides an effective description of the physics at energy
scales for which the fifth dimension cannot be resolved.  However, had
we not used the rescaled cutoff, we would have obtained cutoff
dependent terms that are proportional to $a^2$ in the effective
potential.  No counterterm exists to absorb such terms.  On the other
hand, the rescaled cutoff yields $V\propto a^4$, which is a pure
counterterm that can be absorbed into the TeV brane tension.

We can also see this result using dimensional regularization.  The
dimensionally reduced theory becomes
\begin{equation}
{\cal L}={f_n^{n-2}\over 2} \left(\partial a^{{n-2}\over
2}\right)^2-\delta V_v\mu^{n-4} a^n,
\end{equation}
where $f_n$ has dimensions of mass.  Introducing a canonically
normalized radion field $\varphi=(f_na)^{(n-2)/2},$ the vacuum energy
scales as
\begin{equation}
V\sim \left(\delta V_v\mu^{n-4}\over
f^n_n\right)^{n/2}\varphi^{2n\over n-2}=\left(\delta V_v\mu^{n-4}\over
f^{n-2}_n\right)^{n/2} a^n,
\end{equation}
which simply renormalizes the TeV brane tension.  Finally, one could
also use a Pauli-Villars regulator.  To avoid cutoff-dependent terms
that cannot be absorbed into counterterms, the regulator masses should
scale with $a$ in the same way as masses on the TeV brane.  In this
case, the resulting effective potential is proportional to $a^4$ in
agreement with the two other methods described here.

\section{Conclusion}

In this paper we have calculated the quantum effective potential for the radion in compactified $\mbox{AdS}_5$.  By explicitly performing the computation using three different regulators, we have shown that fields confined to the TeV brane give no non-trivial contributions to the potential.  In particular, we find that in dimensional regularization, the disappearance of any contribution that scales as $a^4\ln a$ is not due to a rescaling of the regulator mass $\mu$ by a factor of $a$.  Instead, it can be traced to the fact the proper generally covariant counterterm in $n=4-\epsilon$ dimensions includes this term.  Likewise, general covariance requires that within a cutoff regularization procedure, one should use the rescaled cutoff $\Lambda a$, leading to $V\propto a^4.$

The contribution due to bulk fields yields a non-trivial dependence on the warp factor $a.$  However, as in the case of confined fields, no terms of the form $a^4\ln a$ are generated.  Beyond the pure counterterm $a^4$, bulk fields generate terms that are suppressed in the large $r_c$ regime.  For instance, a massless bulk field yields terms of the form $a^6$ as well as the finite log term $a^8\ln a.$  This $a$ dependence is too weak to generate an exponentially small value of $a$ without having to choose  unnatural values of the brane tensions.  Because of this, a classical stabilization mechanism is needed.  Our results for the bulk fields disagree with the results of \cite{toms}, who finds enhanced power dependence on $a$, but agree with those of \cite{garriga}. However, our method and interpretations seem to differ from this reference. We have shown how to properly calculate the vacuum energy in the effective theory by first dimensionally reducing and then summing over modes.

Finally, we have also included the quantum effects of the radion field itself.  We found that as in the case of TeV brane fields, the correct momentum space cutoff should be rescaled by a factor of $a$.  While this is quite natural for
brane fields it is not obvious that this had to be so for the radion field, since in the dimensionally reduced theory we have integrated over the fifth dimension, and there is no single preferred ``scale'' $\exp(-kr_c \phi)$. 

\acknowledgments The authors benefited from conversations P. Horava, K. Intrilligator and H. Ooguri.  We thank M. Wise for helping us find an error on an earlier version of this paper.  This work was supported in part by the Department of Energy under grant numbers DE-FG03-92-ER 40701 and DOE-ER-40682-143.

\appendix
\section{Erratum}

The expansion of the expression Eq.~(20) in the small $a$ limit is not valid for the special case in which the formula Eq.~(13) for the Kaluza-Klein masses involves Bessel functions of order zero.  This case corresponds to bulk gauge fields or to scalars and fermions with bulk masses $m^2=-4k^2$ and $m=k/2$, respectively. In these cases the radion effective potential for the contains terms of the form $a^4/\ln a$.  For instance, for a bulk gauge field satisfying Neumann boundary conditions on the two branes Eq.~(20) gives
\begin{equation}
V=V_h+ V_v a^4 + {k^4 a^4\over 16\pi^2}\int_0^\infty dt \, t^3 \ln\left[1-{K_0(t) I_0(a t)\over I_0(t) K_0(a t)}\right],
\end{equation}
which for small $a$ can be approximated by
\begin{equation}
V\simeq V_h + V_v a^4 + {k^4\over 16\pi^2} {a^4\over\ln a}\int_0^\infty dt \, t^3 {K_0(t)\over I_0(t)}
\end{equation}
which has a maximum at $a\simeq \exp\left(-k^4 C_0/16 \pi^2 V_v\right)$, where $V_v$ is the deviation of the TeV brane tension from the background value $-24 M^3 k,$ and $C_0\simeq 1.00453$.  It has been pointed out in~\cite{Garriga:2002vf} that changing the boundary conditions can change the sign of the logarithmic term, yielding a minimum for the radion.  However, even if one could reverse the sign of this calculable quantum correction to the radion potential, one would need to have $V_v\sim k^4/(192\pi^3)$ in order to obtain a radion VEV with the correct magnitude to yield the appropriate hierachy.  Given that $k/M <1$ in order to preserve calculability, a mechanism for radion stabilization that requires such values of $V_v$ should be regarded as fine tuned (for instance using a typical value $M/k\sim 10$, we find that a fine tuning to one part in $10^7$ is necessary).

\end{document}